\documentclass[10pt,twoside]{article}
\usepackage{amssymb}
\usepackage{latexsym}

\makeatletter

\@addtoreset{equation}{section} \makeatother
\newtheorem{theorem}{Theorem}[section]
\newtheorem{remark}[theorem]{Remark}
\newtheorem{lemma}[theorem]{Lemma}
\newtheorem{proposition}[theorem]{Proposition}
\newtheorem{corollary}[theorem]{Corollary}
\newtheorem{example}[theorem]{Example}
\newtheorem{definition}[theorem]{Definition}

\newcommand{\bdefi}{\begin{definition}}
\newcommand{\btheo}{\begin{theorem}}
\newcommand{\bprop}{\begin{proposition}}
\newcommand{\brema}{\begin{remark}}
\newcommand{\bcoro}{\begin{corollary}}
\newcommand{\blemm}{\begin{lemma}}
\newcommand{\bexam}{\begin{example}}

\newcommand{\edefi}{\end{definition}}
\newcommand{\etheo}{\end{theorem}}
\newcommand{\eprop}{\end{proposition}}
\newcommand{\erema}{\end{remark}}
\newcommand{\ecoro}{\end{corollary}}
\newcommand{\elemm}{\end{lemma}}
\newcommand{\eexam}{\end{example}}

\newcommand{\be}{\begin{equation}}
\newcommand{\ee}{\end{equation}}

\newcommand{\R}{{\mathbb R}}

\newcommand{\ben}{\begin{enumerate}}
\newcommand{\een}{\end{enumerate}}
\newcommand{\bit}{\begin{itemize}}
\newcommand{\eit}{\end{itemize}}
\newcommand{\edoc}{\end{document}}

\newcommand{\mm}{S}

\title{
\vspace{0.5in} {\bf Physics from scratch.} \\ {\em Letter on M.
Tegmark's ``The Mathematical Universe'' }}

\author {\bf Antonio N. Bernal$^{1}$,
Miguel
S\'anchez$^{1}$, Francisco Jos\'e Soler Gil$^{2}$
\\
$^1$ {\it\small Departamento de Geometr\'{\i}a y Topolog\'{\i}a}\\
{\it\small 
Universidad de Granada, Spain}\\
$^2$  {\it\small Institut f\"ur Philosophie} \\
{\it\small Universit\"at Bremen, Germany}\\
{\small sanchezm@ugr.es, soler@uni-bremen.de}}

\begin{document}

\parindent=5mm
\date{}
\maketitle

\begin{quote}

\noindent {\small \bf Abstract.} {\small In a recent article, M.
Tegmark \cite{Te} poses the hypothesis that our known universe is
a ``baggage free'' mathematical structure among many other
possible ones, which also correspond to other physical universes
---Mathematical Universe Hypothesis, MUH. Naturally, questions
arise, such as how to obtain the  physical properties of our world
from the mathematical structure, or how many possibilities exist
for a Universe minimally similar to ours.

In this letter we present some results which can be regarded as a
strengthening of MUH, as they give some hints on the derivation of
spacetime in current physics from a baggage free mathematical
structure.

Concretely, we argue that the set of mathematical structures which
can be interpreted as a description of a spacetime is drastically
reduced, if one admits some natural postulates on minimal
symmetry. Furthermore, the apparently very particular form of
classical Galilei-Newton and relativistic spacetimes, is not
arbitrary and cannot be regarded as ``two possibilities among
arbitrarily many others''. In fact, such theories are determined
by a single mathematical structure which only permits four
possible types of spacetimes. Finally, we show how the minimal
postulates on symmetry can be endowed with a simple physical
interpretation, i.e., they acquire ``baggage'' in a natural way.
}\\

\end{quote}



\newpage


\section{Introduction}

In his recent article \cite{Te} (see also \cite{Te0}), Tegmark
explored the implications of the {\em Mathematical Universe
Hypothesis} (MUH) which, essentially, claims the full
identification of both notions, ``mathematical existence'' and
``physical existence''. According to MUH, our universe is really a
mathematical structure, and conversely: each consistent
mathematical structure would be also a physically real universe.
With this proposal, Tegmark argues the possibility to answer some
of the deepest questions about the foundations of physics, such as
Wigner's {\em why are so effective mathematics in natural
sciences?} or Wheeler's {\em why these equations and no others?}
Tegmark's picture of the world is then a physical reality with no
free parameters and no arbitrary elements of any kind: all
non-contradictory possibilities are real. Consequently, nothing is
contingent, and no open question remain.

From the very beginning, the authors of the present article must
recognize that they are not very enthusiastic about these ideas.
For the simple reason that Tegmark's proposal leaves important
elements unexplained such as, for example,  the elusive reality of
the time (how to do justice to the details of our experience of
the ``now'' and the temporal becoming of reality in the context of
a tenseless mathematical world?). Certainly, such troubles are
shared by our major physical theories ---General Relativity, in
the case of time. But most people makes no claims in order to
identify these theories with the totality of physical reality.

Nevertheless, Tegmark's proposal is very attractive, and worth
exploring. Therefore, in the next pages, we develop some ideas
which could support MUH. They rely on a postulational basis for
spacetime introduced  from a  different viewpoint  by two of the
authors (ANB \& MS) and  M. L\'opez \cite{BLS}. Concretely:

\ben \item We postulate that the elements of a certain
mathematical structure, with no physical baggage, satisfy certain
symmetry (Section \ref{s2}). As a consequence, four and only four
variants of this structure can exist (Section \ref{s3}). \item Two
of these variants correspond to a generalization of the
Galilei-Newton theory and General Relativity (Section \ref{s4}).
This stresses: \bit \item The concrete form of the physical laws
of our universe can be determined by symmetry conditions on a
purely mathematical underlying structure. \item To postulate a
minimum of symmetry in such an underlying mathematical structure,
 drastically limits the number of possible physical scenarios. \eit
\item The conditions of symmetry and other aspects can be easily
endowed with physical meaning (Section \ref{s5}). This suggests
that many properties perceived by physical observers may emerge
from the mathematical properties of the underlying structure. \een

 For simplicity, we focus on the
structure of classic spacetime. In the case that our universe were
a mathematical structure, such a structure would  also be
responsible for the quantum aspects of the world. Nevertheless, to
consider the classic case is enough, in order to have a taste of
the
 dependence   of many particularities of the
physical world on purely mathematical features. This is precisely
the point which makes MUH so suggestive and attractive.

\section{A nice mathematical structure} \label{s2}

Consider the following simple mathematical background structure
(MB):
\begin{quote}
{\bf MB}. The pair $(M, \mm)$, where $M$ is a smooth connected
4-manifold, with differentiable structure denoted as ${\cal D}$,
and $\mm: M \rightarrow {\cal P}({\cal D}), p\mapsto \mm_p$ is a
map (with codomain the set of parts of\footnote{i.e., ${\cal D}$
is the set of all the smooth coordinate charts of $M$ and  ${\cal
P}({\cal D})$ is the set of all the subsets of ${\cal D}$.} ${\cal
D}$) which satisfies:

$\mm_p$ is non-empty, and all the charts in $\mm_p$ are defined on
some neighborhood of $p$.
\end{quote}
Notice that each element of ${\cal D}$ is a coordinate chart
$(U,\phi), \phi: U\subseteq M \rightarrow \R^4$. The (ordered)
coordinates will be denoted as $(x^0,x^1,x^2,x^3)$. Now, let us
enrich MB with a symmetry postulate (SP):

\begin{quote}
{\bf SP}. Any two charts $O\equiv (x^0,x^1,x^2,x^3)$, $\tilde
O\equiv (\tilde x^0,\tilde x^1,\tilde x^2,\tilde x^3)$ in $\mm_p$
satisfy:
\begin{equation} \label{ep2}
\partial_{\,x^0}\,{\tilde x^0}|_p=\partial_{\,{\tilde x^0}}\,x^0|_p \; , \,\;\,\,\;\,
 \partial_{\,x^j}{\tilde x}^i|_p=\partial_{\,{\tilde x}^i}x^j|_p \; ,\,\;\forall i,j\in\{1,2,3\}.
\end{equation}
\end{quote}
Notice that this hypothesis  assumes two independent symmetries
between the different charts selected for each $p$: a symmetry for
the 0-th coordinate and a symmetry for the set of the other three
coordinates.

Finally, we will restrict our attention to the {\em smooth class}
{\bf SC} of mathematical structures where the symmetry assumption
SP fits smoothly with the ambient manifold $M$. The precise
formulation of this class becomes trivial when the consequences of
SP are analyzed, as we will see below.

\section{The set of resulting mathematical structures} \label{s3}

Previous elements MB, SP and SC
correspond to  postulates in an approach 
for the study of spacetime  in \cite{BLS}. This approach develops
a different viewpoint (in Tegmark's terms, from the frog to the
bird viewpoint\footnote{See also \cite{Sa} for additional
developments which may be useful in this section.}), but its
careful mathematical analysis is applicable now.

Let $S^1$ be the circumference obtained from the extended real
line $[-\infty, \infty]$ by collapsing $\pm \infty$ to a single
point $\omega$. As proven in full detail in \cite{BLS}:

\begin{quote}{\em
 Under the background MB, the symmetry postulate SP allows to attach
 to the tangent space $T_pM$ at $p$ both, some $k(p)\in S^1$ and one among four concrete linear
 structures.}
\end{quote}
More precisely, these four structures are:

\ben \item If $k(p)\in (-\infty,0)$: a Lorentzian scalar product
on $T_pM$. (The value of $k(p)$ corresponds to the normalization
of $\partial_{x_0}|_p$ for all the charts in $\mm_p$.) \item If
$k(p)=\omega$: a (non-zero) linear form $\Omega_p\in T_pM^*$
($\Omega_p: T_pM \rightarrow \R$) and an Euclidean scalar product
$h_p$ on the kernel of $\Omega_p$. \item If $k(p)=0$: a (non-zero)
tangent vector  $Z_p\in T_pM$ and an Euclidean scalar product
$h^*_p$ on the kernel of $Z_p$ in $T_pM^*$.  \item If $k(p)\in
(0,\infty)$: an Euclidean scalar product on  $T_pM$. \een

As explained in \cite{BLS}, the restriction to the smooth class SC
consists simply in the restriction to the case where the function
$k(p)$ (and the corresponding linear structure), vary smoothly
with $p$.

So far, we have considered only a purely mathematical structure
or, in Tegmark's terminology, a bird viewpoint of (a piece of)
reality. Next, we will descend from this Platonic world, looking
for the physical meaning of the mathematical  structure.

\section{The (very small) family of classical spacetimes } \label{s4}

The a priori deduction of the physical properties of a
mathematical structure, is a serious challenge for defenders of
MUH. Tegmark poses the problem as follows \cite[Section III]{Te}:

\begin{quote} {\em
Suppose we were given mathematical equations that completely
describe the physical world, including us, but with no hints about
to interpret them. What would we do with them? Specifically, what
mathematical analysis of them would reveal their phenomenology,
i.e., the properties of the world that they describe as perceived
by observers? [...] This is in my opinion one of the most
important questions facing theoretical physics, because if we
cannot answer it, then we cannot test a candidate TOE by
confronting it with observation. [...] By construction, the only
tools at our disposal are pure mathematical ones, so the only way
in which familiar notions and interpretations (``baggage") can
emerge are as implicit properties of the structure itself that
reveal themselves during the mathematical investigation.
}\end{quote}

Clearly, the most rigorous way to descend from the bird's
viewpoint to that of the frog (the viewpoint of an observer who
lives in the world and is part of it), is to start from the purely
mathematical analysis and then to detect the emergency of the
physical concepts
---as implicit properties of the analyzed structures.
Nevertheless, this ideal procedure is not only difficult but also
vague, as  the meaning of ``a physical property implicit in a
mathematical structure'' is by no means clear.

Thus, instead of following this path, our starting point in this
section will be the already known physics. The comparison between
the previously introduced mathematical structures and the known
theories of spacetime is trivial. However, as we will see, not all
the results of the comparison are trivial. Some of them can be
interpreted as a strengthening of the intuitions underlying  MUH
---a step towards understanding the emergency of physical properties from
certain aspects of mathematical structures.

Let us start with the most obvious one: \ben \item Structures with
$k(p)\in (-\infty,0)$, for all $p\in M$. Recall that in this case
a scalar Lorentzian product at each tangent space is obtained.
That is, one has a Lorentzian metric on a 4-manifold ---the
ambient for spacetime of classical General Relativity.
Additionally, all the orthonormal basis at each $T_pM$ can be
chosen normalized so that the metric on the first vector is equal
to $k(p)$. Putting $c(p)^2=-k(p)$ a closer analogy to General
Relativity is obtained. \item  $k \equiv \omega$. One has a
non-vanishing 1-form $\Omega$ on $M$ and a Riemannian metric $h$
on the fiber bundle obtained as the kernel of $\Omega$. This is
called a {\em Leibnizian structure} in \cite{BLS}, and it is
studied exhaustively in \cite{BS}. It is a clear generalization of
Galilei-Newton spacetime, where  $\Omega$ is an exact form (the
differential $\Omega=dT$ of the ``absolute time'' $T$), and a flat
connection $\nabla$ which parallelizes $h$ is assumed to exist.

\item   $k\equiv 0$. This  case is mathematically dual and
analogous to the previous one, even though it does not have
classical analog.

\item $k(p)\in (0,\infty)$, for all $p\in M$. Reasoning as in the
first case, now one has a (positive-definite) Riemannian metric on
the manifold $M$. Notice that, according to SP, one only assumes
the existence of independent symmetries between, on one hand,  the
first coordinates, and, on the other, the sets composed of the
other three coordinates  (formula (\ref{ep2})). This does not,
however, exclude the existence of additional symmetries between
all the coordinates ---which is the present case. There are slight
 strengthenings  of SP which would forbid this case (say,
assuming that the members of the first equality in (\ref{ep2}) are
positive). Nevertheless, according to MUH, our more general
version of SP must be also taken into account ---and suggests
interesting possibilities.

\een Now, let us emphasize a first interesting conclusion: the
relation between the Galilei-Newton and relativistic structures of
spacetime. They are singled out as members of a family of only
four elements ---the mathematical structures compatible with MA,
SP, SC. This suggests that classical and relativistic physics are
not two arbitrary physical scenarios within a set of innumerable
alternatives.

Considering the most radical possibility, the introduced
mathematical structure might be the bases from which classical
spacetimes arise. Of course, there are more general possibilities:
consider, for example, the case of $n$ coordinates and independent
symmetries for the first $m<n$ ones. Nevertheless, these scenarios
may be so different that  they might not contain observers
---say, self-aware structures in a sense minimally similar to us.
So, Wheeler's question in this context (why these spacetimes and
no others?) would admit a first answer: there were not so many
possible alternatives. To which extent this conclusion is sound
must be examined carefully but, in any case, it is clearly worth
taking into account.

Assuming the conclusion as a working hypothesis, the following
appealing item is the key role played by the symmetry condition
SP. Essentially, this symmetry seems to be so powerful that it
generates all the spacetimes. This seems to agree with Tegmark's
claim:

\begin{quote}
{\em [...] when studying a mathematical structure S to derive its
physical phenomenology (the ``inside view''), a useful first step
is finding its symmetries [...]}
\end{quote}
Thus, next we will study next the physical meaning of our
spacetime-generating condition of symmetry ---beyond a purely
mathematical element.

\section{The physical  meaning of the  symmetry condition} \label{s5}

Let us reformulate PS as follows:

\begin{quote}
For each point $p\in M$ there exists a non-empty subset $S_p$ of
coordinate charts or {\em spacetime observations around } $p$,
the first coordinate $t\equiv x^0$ of each chart
  labelled as the {\em time coordinate} and the other three $x^i, i=1,2,3$ as the {\em spatial
 coordinates}. Then, given two such charts
$O\equiv (t,x^1,x^2,x^3), \tilde O\equiv (\tilde t,\tilde
x^1,\tilde x^2,\tilde x^3)$ in $\mm_p$ there exist a symmetry at
$p$ between the timelike coordinates:
\begin{equation} \label{epa}
\partial_{\,t}\,{\tilde t}|_p=\partial_{\,{\tilde t}}\,t|_p
\end{equation}
and also between the spatial coordinates:
\begin{equation} \label{epb}
 \partial_{\,x^j}{\tilde x}^i|_p=\partial_{\,{\tilde x}^i}x^j|_p \; ,\,\;\forall i,j\in\{1,2,3\}.
\end{equation}
\end{quote}
This condition was introduced above as a mere element of the
mathematical structure. Nevertheless, as the names introduced now
suggest (and  proved in full detail in \cite{BLS}), such an
element can be endowed with a simple physical meaning. Summing up:
\bit \item The set $S_p$ of  privileged charts is just the set of
observations carried out by the {\em standard observers} of the
theory (say, at least infinitesimally standard observers at $p$:
inertial, freely falling, using ideal instruments, selected by
some sort of ether...) \item The symmetry (\ref{epa}) between the
timelike coordinates ensures that the measures of time cannot
privilege one of the two standard observers over the other one.
That is, the ${\tilde O}$-time, measured with the $O$ clock, goes
by as the $O$-time, measured with the ${\tilde O}$ clock ---$O$
and ${\tilde O}$ measure at $p$ using the same ``fundamental unit
of time''. \item Analogously, the second symmetry (\ref{epb})
ensures the inexistence of privileges for the spatial coordinates
or, equally, the use of the same fundamental space units by $O$
and $ \tilde O$.
 \eit
Summing up, the maximum number of classic spacetime theories seem
to be just four, and: (a) this is obtained  from a symmetry
mathematical postulate on a 4-manifold, and (b) from this
postulate, a simple baggage interpretation emerges. These two
facts agree with Tegmark's picture derived from MUH.

\section{Conclusion} \label{s6}

We repeat that the authors do not agree with the full
identification of both, reality and mathematical structures, which
lies in the core of MUH. Long discussions on this topic are
possible, and we have already mentioned the problem of our
experience of time
---real or illusory, in any case unexplained\footnote{See
\cite{Cr, Soler} about the difficulties of explaining the human
experience of time starting at the ``tenseless'' existence of the
reality.}. This is only  an example of our reluctance among
others: the unexplained experience of subjective perceptions
(``qualia''), the apparent superfluous character of quantum level
3 multiverses, the unclear falsability\footnote{One of the authors
(FJSG) thinks that there could be in principle uncountably many
universes similar to ours, obtained for example by allowing the
``constants of nature" to be variable. If this were true, the
generic classic universes would be those with ``almost constants"
of nature (...or they could be universes still more irregular,
perhaps). As such almost constants are not found, we would be
living in a very particular universe. And this would not fit well
in MUH. A possible answer would be that even a minimal change in
the constants of the world precludes the existence of observers.
But, in this case, is MUH testable at all?} of the theory...
However, to dispute  this topic is not our purpose.

Anyway, we have seen along the present letter that, at least in
the case of classic spacetimes, the relation between the
mathematical formalism  and the physical meaning fits  the
expectations of MUH: starting at the symmetries of a concrete
mathematical structure, the physical properties perceived by the
(frog) observers emerges.

This reinforces MUH. Perhaps we would not bet that the universe is
as Pythagorean as Tegmark suggests, but the existence of a strong
Pythagorean component is undisputable.

\section*{Acknowledgments}

MS has been partially supported by Spanish MEC-FEDER Grant
MTM2007-60731 and Regional J. Andaluc\'{\i}a Grant P06-FQM-01951.

\end{document}